\documentclass[twocolumn,tighten]{aastex631}

\newcommand{\ltsima}{$\; \buildrel < \over \sim \;$}
\newcommand{\simlt}{\lower.5ex\hbox{\ltsima}}

\def\arcdeg{\hbox{$^\circ$}}
\def\arcmin{\hbox{$^\prime$}}
\def\arcsec{\hbox{$^{\prime\prime}$}}

\usepackage{amsmath}

\shorttitle{Hot Gas Outflows of NGC~4945}
\shortauthors{Porraz Barrera et al.}

\begin{document}

\title{Hot Gas Outflow Properties of the Starburst Galaxy NGC~4945}

\author[0009-0001-0715-7209]{Natalia Porraz Barrera}
\affiliation{Department of Astronomy, The Ohio State University, 140 W. 18th Ave., Columbus, OH 43210, USA}

\author[0000-0002-2644-0077]{Sebastian Lopez}
\affiliation{Department of Astronomy, The Ohio State University, 140 W. 18th Ave., Columbus, OH 43210, USA}
\affil{Center for Cosmology and AstroParticle Physics, The Ohio State University, 191 W. Woodruff Ave., Columbus, OH 43210, USA}

\author[0000-0002-1790-3148]{Laura A. Lopez}
\affiliation{Department of Astronomy, The Ohio State University, 140 W. 18th Ave., Columbus, OH 43210, USA}
\affil{Center for Cosmology and AstroParticle Physics, The Ohio State University, 191 W. Woodruff Ave., Columbus, OH 43210, USA}

\author[0000-0002-1616-1701]{Adi Foord}
\affiliation{Department of Physics, University of Maryland Baltimore County, 1000 Hilltop Circle, Baltimore, MD 21250, USA}

\author[0000-0002-1875-6522]{Dustin D. Nguyen}
\affil{Center for Cosmology and AstroParticle Physics, The Ohio State University, 191 W. Woodruff Ave., Columbus, OH 43210, USA}
\affil{Department of Physics, The Ohio State University, 191 W. Woodruff Ave, Columbus, OH 43210, USA}

\author[0000-0003-2377-9574]{Todd A. Thompson}
\affil{Department of Astronomy, The Ohio State University, 140 W. 18th Ave., Columbus, OH 43210, USA}
\affil{Center for Cosmology and AstroParticle Physics, The Ohio State University, 191 W. Woodruff Ave., Columbus, OH 43210, USA}

\author[0000-0003-2377-9574]{Smita Mathur}
\affil{Department of Astronomy, The Ohio State University, 140 W. 18th Ave., Columbus, OH 43210, USA}
\affil{Center for Cosmology and AstroParticle Physics, The Ohio State University, 191 W. Woodruff Ave., Columbus, OH 43210, USA}

\author[0000-0002-5480-5686]{Alberto D. Bolatto} 
\affil{Department of Astronomy, University of Maryland, College Park, MD 20742, USA}
\affil{Joint Space-Science Institute, University of Maryland, College Park, MD 20742, USA}

\begin{abstract}

We analyze 330 ks of {\it Chandra} X-ray imaging and spectra of the nearby, edge-on starburst and Seyfert Type 2 galaxy NGC~4945 to measure the hot gas properties along the galactic outflows. We extract and model spectra from 15 regions extending from $-$0.55 kpc to $+$0.85 kpc above and below the galactic disk to determine the best-fit parameters and metal abundances. We find that the hot gas temperatures and number densities peak in the central regions and decrease along the outflows. These profiles are inconsistent with a spherical, adiabatically-expanding wind model, suggesting the need to include mass loading and/or a non-spherical outflow geometry. We estimate the mass outflow rate of the hot wind to be $1.6\:M_{\odot}~\rm{yr}^{-1}$. Emission from charge exchange is detected in the northern outflow, and we estimate it contributes 12\% to the emitted, broad-band ($0.5-7$~keV) X-ray flux.

\end{abstract}


\section{Introduction} \label{sec:intro}

The presence of galactic outflows in star-forming galaxies is well established. These star formation feedback driven outflows have important effects on their host galaxies such as distributing metals to the circumgalactic and intergalactic medium, quenching star formation, and affecting the metallicity of the galactic disk \citep{Tumlinson2017,Mathur2022}. These outflows, also known as galactic winds, are multiphase and can be observed in various wavelengths, each with their own set of constraints \citep{Veilleux2005,Veilleux2020}. 

NGC~4945 is a nearby ($D = 3.72$~Mpc; \citealt{Tully2016}) edge-on ($i=78^{\circ}$; \citealt{Ott2001}) galaxy that lies in the Centaurus A/M83 group and is known to host a central starburst \citep{Schurch2002,Emig_2020} as well as a Seyfert Type 2 AGN \citep{Iwasawa1993}. The mass of the central black hole is $M_{\rm BH} = 1.4 \times {10^6}\:M_{\odot}$, and the AGN accretes at 10 $-$ 30\% of the Eddington rate \citep{Puccetti14}.

Similar to M82 and NGC~253, NGC~4945 is one of the nearest far-IR galaxies (FIRGs; \citealt{Heckman1990}) producing large-scale outflows along the galaxy's minor axis. The outflow has been observed at mm \citep{Bolatto2021}, optical \citep{Venturi2017,Mingozzi2019}, and X-ray wavelengths \citep{Schurch2002,Done2003, Strickland20041, Strickland20042,Marinucci2012, Marinucci2017}. In this paper, we focus on {\it Chandra} X-ray observations, constraining the temperature and metallicity gradients of the hot gas ($\sim10^{7}$~K) in the outflow.

Previous work on the hot phase of NGC~4945's outflows constrained the temperatures as well as the properties of the central AGN \citep{Schurch2002,Done2003}. Using {\it XMM-Newton} and {\it Chandra} observations, \cite{Schurch2002} extracted spectra to identify the emission lines of the outflow and nuclear central region. The central region was fitted with a Compton-reflection component, an iron line, and a hot thermal component, while the outflow region was fitted with three thermal components. \cite{Done2003} measured column densities and temperatures in four regions (one from the nuclear source and three along the outflow) by fitting two temperature MEKAL plasma components in the diffuse regions and an iron line and power law in the nuclear region using {\it Chandra} and {\it RXTE} data. \cite{Marinucci2017} analyzed the circumnuclear environment of the galaxy to understand the central structure of the source using {\it Chandra} observations. Spectra were extracted from five circular circumnuclear regions and were fitted with a model that consisted of a reflection continuum and five emission lines to obtain the energy, flux, and equivalent width values for each of the regions.

In this paper we follow the analysis of \cite{Lopez2020,Lopez2023} to constrain temperature, density, and metal abundances of the outflow in NGC~4945 using 330 ks of archival {\it Chandra} data. Previous works have not constrained the metal abundances in the wind nor have they tested the significance and contribution of charge-exchange (CX) to the total X-ray emission in NGC~4945. In CX, ions strip electrons from neutral atoms; in galactic winds, CX occurs as the hot phase interacts with the cooler gas. Previous works on other nearby starbursts M82 and NGC~253 have found that CX can contribute significantly to the total broad-band ($0.5 - 7$ keV) flux, altering the metal abundance estimates substantially if CX is not considered \citep{Zhang2014,Lopez2020,Lopez2023}.  

This paper is structured as follows. In Section~\ref{sec:methods}, we describe the archival {\it Chandra} observations and the procedure to create X-ray images and to extract and model spectra. In Section~\ref{sec:results}, we present temperature, density, and metallicity profiles as well as geometric constraints on the outflows. In Section~\ref{sec:discussion}, we compare our results to previous work on NGC~4945 (Section~\ref{sec:previouswork_NGC4945}) and on other nearby galaxies with starburst-driven outflows, M82 and NGC~253 (Section~\ref{sec:previouswork_starbursts}). We also show how the observed temperature and density profiles in NGC~4945 contrast the predictions of a spherically symmetric, adiabatically expanding wind (Section~\ref{sec:windmodels}). We measure the hot gas outflow rates in Section~\ref{sec:massoutflowrates}, and we quantify the contribution of charge exchange emission to the emitted X-ray flux in Section~\ref{sec:CX}. In Section~\ref{sec:conclusions}, we summarize our findings and outline future work.

Throughout the paper we assume a distance of 3.72 Mpc to NGC~4945 \citep{Tully2016} and a redshift of $z = 0.001878$ \citep{Allison2014}. 

\section{Methods}\label{sec:methods}
\subsection{Observations and Data Reduction}

\begin{figure*}
    \centering
\includegraphics[width=0.95\textwidth]{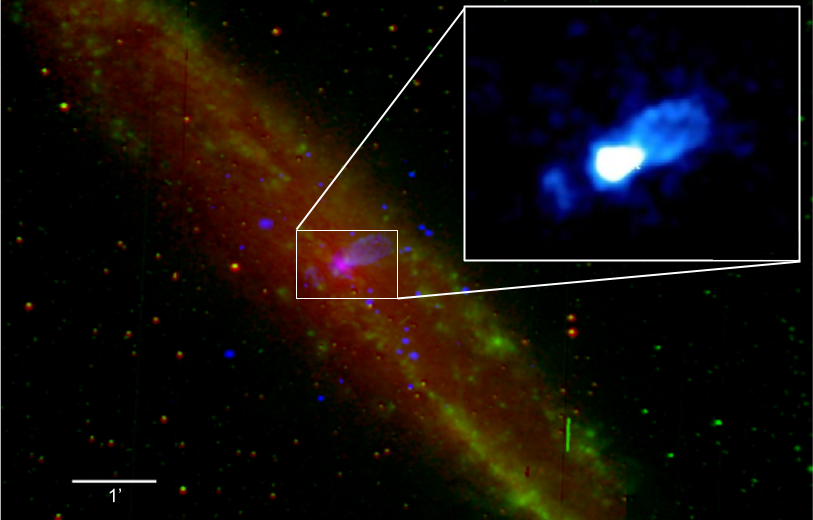}
    \caption{Three color image of NGC~4945. Blue is the 0.5-7.0 keV {\it Chandra} X-rays, green is H$\alpha$ \citep{Halpha}, and red is 2.2 $\mu$m infrared \citep{infrared} from \cite{ned1}. North is up, and East is left. Zoomed-in picture (top right) shows the diffuse, hot gas of the outflows in X-rays with the point sources removed. At the distance of NGC~4945 1\arcmin{} is about 1 kpc.}
    \label{fig:3colorimage}
\end{figure*}

NGC~4945 has been observed six times by {\it Chandra} with the Advanced CCD Imaging Spectrometer (ACIS) from 2000 to 2018 (see Table~\ref{table:data}). After removing 20 ks of exposure time from the first observation (ObsID 864) due to background flaring (times when the count rate surpassed 10 count~s$^{-1}$), we obtained a total exposure time of $\approx$330 ks. Our data reduction was performed using version 4.14 of the Chandra Interactive Analysis of Observations {\sc ciao} \citep{CIAO}.  

With the {\sc ciao} command \textit{merge\_obs}, the six observations were merged into a single broad-band (0.5 $-$ 7.0 keV), exposure-corrected X-ray image, and with the \textit{wavdetect} command, point sources were identified. These point sources were then removed with the \textit{dmfilth} command to construct an image of the diffuse gas associated with NGC~4945's hot outflows. 

Figure~\ref{fig:3colorimage} shows a 3-color image of the galaxy. The green and red colors are the galacti1c disk in H$\alpha$ \citep{Halpha} and the 2.2 $\mu$m infrared \citep{infrared} respectively, and the blue is the diffuse hot gas in broad-band (0.5 $-$ 7.0 keV) X-rays, which extends $\sim$0.85 kpc north and $\sim$0.55 kpc south from the galaxy's disk. In the galaxy-wide image, X-ray point sources are present, while in the zoomed-in X-ray only image, these point sources were removed to enable mapping of the diffuse outflows. The gap seen in the southern outflow below the starburst ridge may be due to absorption from the disk along the line-of-sight \citep{Ott2001}. 

\begin{figure}
    \centering
    \includegraphics[width=\columnwidth]{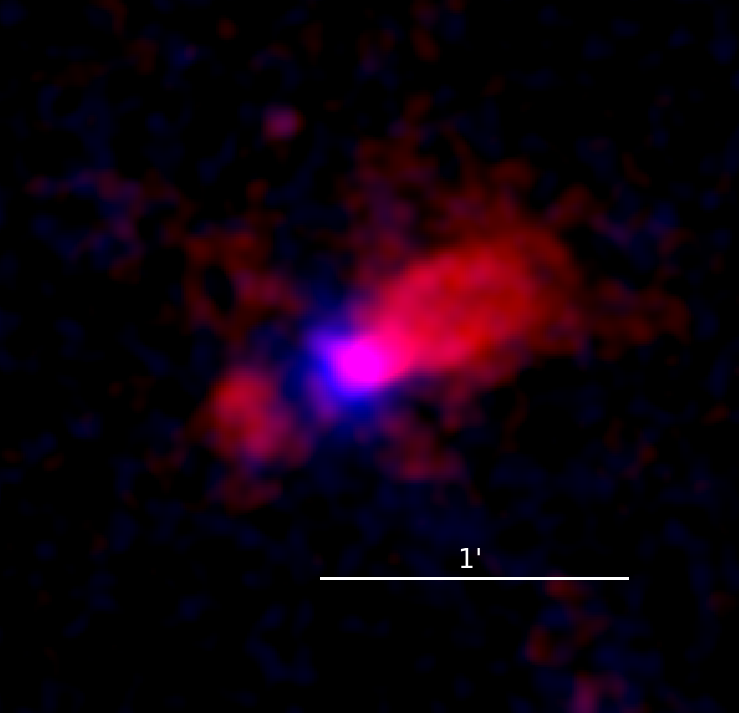}
    \caption{Two-color X-ray image of the soft X-rays (0.5 $-$ 2.0 keV) in red and the hard X-rays (2.0 $-$ 7.0 keV) in blue. Hard X-rays are concentrated in the center due to the AGN and starburst activity. The soft X-rays are more extended, trace the outflow, and are absent in the heavily obscured gap region.}
    \label{fig:2color}
\end{figure}

In Figure~\ref{fig:2color}, we show a two-color X-ray image of NGC~4945, with soft (0.5 $-$ 2.0 keV) X-rays in red and hard (2.0 $-$ 7.0~keV) X-rays in blue. The hard X-rays peak in the center of the galaxy due to the presence of an AGN and the hot gas produced by the starburst. By comparison, the soft X-rays are more extended and trace the material outflowing from the disk.

\begin{deluxetable}{lrc}
\centering
\tablecolumns{3}
\tablewidth{0pt} \tablecaption{{\it 
Chandra} Observations \label{table:data}}
\tablehead{\colhead{ObsID} & \colhead{Exposure (ks)\tablenotemark{a}} & \colhead{Start Date}}
\startdata
864 &   49.12 & 2000-01-27 \\
13791 &	39.45 &	2012-01-10 \\
14412 &	39.14 &	2012-04-03 \\
14984 & 128.76 & 2013-04-25 \\
14985 &	68.73 &	2013-04-20 \\
20997 &	20.71 &	2018-04-24 \\ 
\enddata  
\tablenotetext{a}{Exposure time on ObsID 864 is before 20~ks were removed for background flaring.}
\end{deluxetable}

\subsection{Spectral Analysis}\label{specanalysis}

\begin{figure*}
    \centering    
    \includegraphics[width=0.38\textwidth,trim={0 -49 0 0}]{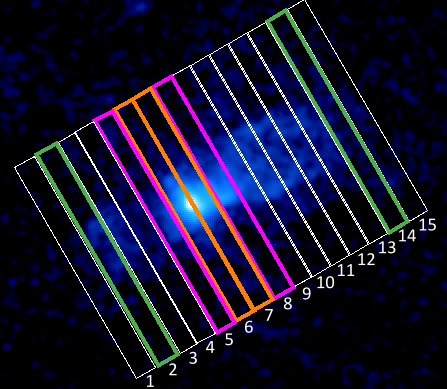} \includegraphics[width=0.6\textwidth]{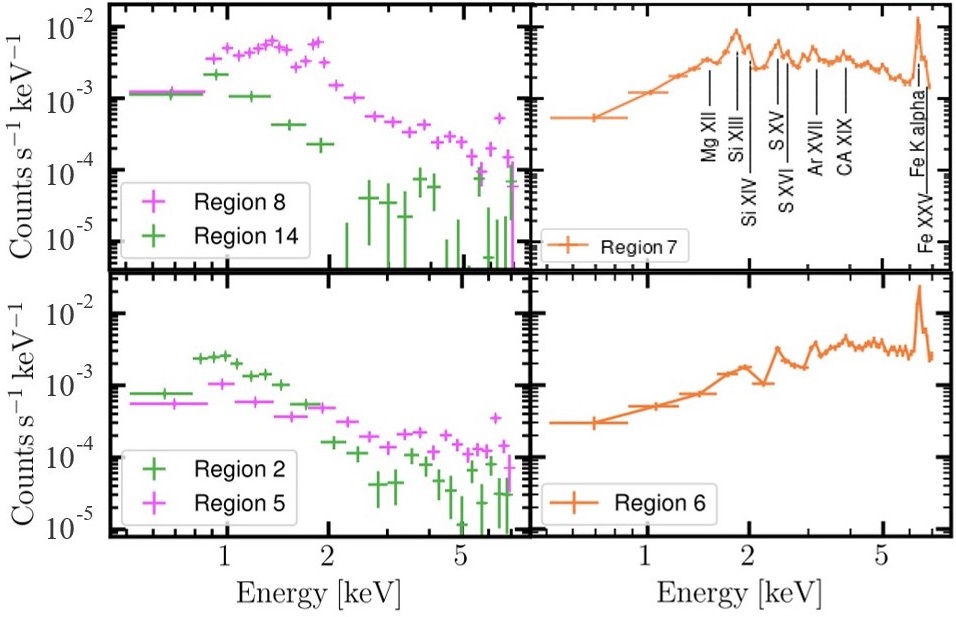}
    \caption{ {\it Left}: Broad-band X-ray image of the NGC~4945 outflows with 15 regions overplotted where spectra were extracted. These regions are 1.0\arcmin\ in height and 0.095\arcmin\ in width. At the distance of NGC~4945 1\arcmin{} is about 1 kpc. {\it Right}: Background-substracted spectra for 6 of the 15 regions. The top (bottom) row corresponds to the spectra from the northern (southern) regions, 8 and 14 (2 and 5), of the outflow. Emission lines from metals are apparent in the central regions, 6 and 7 (orange), including an Fe K$\alpha$ line at 6.4 keV and Fe {\sc xxv} at 6.67 keV. Regions 4 and 5 are located where a ``gap" in the southern X-ray outflow is apparent that is coincident with an enhancement in dense gas (see Figure~\ref{fig:3col_gap}) that likely attenuates the X-rays.}
    \label{fig:spectraplots}
\end{figure*}

\begin{figure}
    \centering \includegraphics[width=\columnwidth]{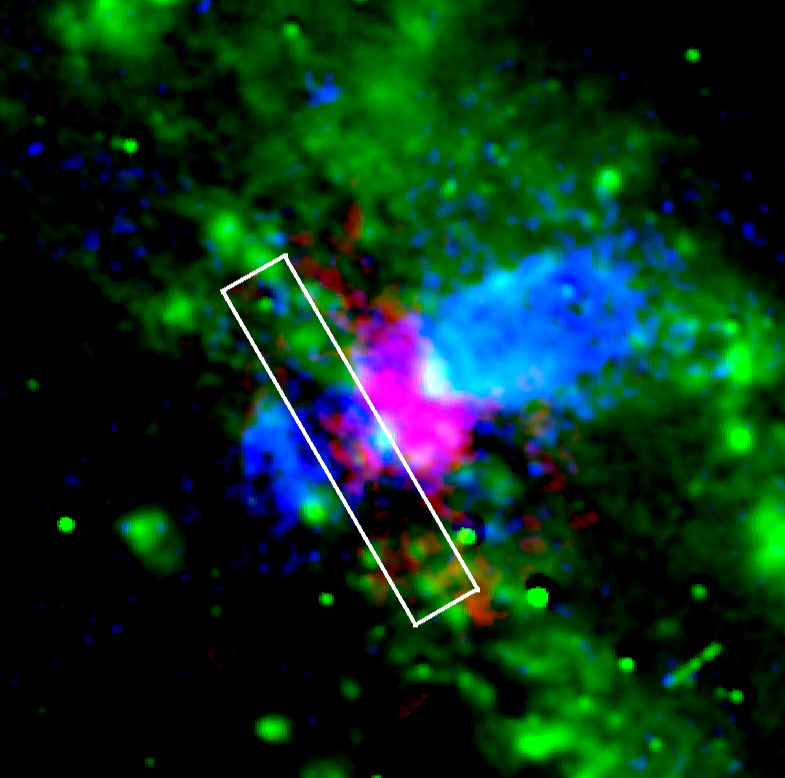}
    \caption{Three-color image of NGC~4945: blue is 0.5 $-$ 7.0 keV {\it Chandra} X-rays, green is H$\alpha$ \citep{Halpha}, and red is CO (2-1) molecular gas (private communication, A. Bolatto). North is up, and East is left. The white box is 1.0\arcmin\ in height and 0.19\arcmin{} in width and represents the combined regions 4 and 5 where spectra were extracted (see Figure~\ref{fig:spectraplots}). At the distance of NGC~4945 1\arcmin{} is about 1 kpc. The gap in X-rays from the southern outflow is coincident with an enhancement of CO (2-1) emission, consistent with the increased column density along the line of sight there.}
    \label{fig:3col_gap}
\end{figure}

To constrain the properties of the hot gas, we extracted spectra using the CIAO command \textit{specextract} from several regions defined along the minor axis of the outflow. The data were grouped with a minimum of ten counts per energy bin. There are 15 total regions, as shown in Figure~\ref{fig:spectraplots}: 5 southern, 2 central, and 8 northern regions, with dimensions 1.0\arcmin\ $\times$ 0.095\arcmin\ and areas of 343 $\mathrm{arcsec}^2$. These dimensions were set in order to achieve at least 500 net counts per region, which allowed us to constrain the electron temperature $kT$ and density $n_{\rm e}$. After examining the spectra, we opted to combine regions 4 and 5 (called ``regions 4 \& 5" hereafter) to increase signal as a ``gap" in the X-rays appears there, likely due to absorption from dense gas along the line of sight, consistent with the presence of CO (2-1) emission in that region (the red in the three-color image in Figure~\ref{fig:3col_gap}). 

The spectra were modeled with XSPEC Version 12.12 \citep{XSPEC}. We adopted solar abundances from \cite{Asplund2009} and photoionization cross sections from \cite{Verner1996}. The XSPEC models used to fit the spectral data varied by region, particularly due to the presence of the AGN. A multiplicative factor ({\sc const}) was included in each region to account for changes in flux/emission measure between the observations. Two absorption components ({\sc phabs*phabs}) per region were also included\footnote{We note that we also performed spectral fitting using two XSPEC absorption component tbabs for the Milky Way and NGC~4945 column densities assuming \cite{wilm00} solar abundances. We found that the spectral results of all regions were statistically consistent within the errors of those reported in Table~\ref{table:values}.} the first component accounts for the Galactic absorption in the direction of NGC~4945 (frozen to the value of $N_{\rm H}^{\rm MW}=1.38\times10^{21}$~cm$^{-2}$; \citealt{MWColumn}), and the second absorption component accounts for NGC~4945's intrinsic absorption $N_{\rm H}^{\rm NGC4945}$ and was allowed to vary. Models from all regions included at least one optically-thin, thermal plasma component ({\sc apec}) and a power-law component ({\sc powerlaw}).

For three southern and eight northern regions (regions 1 $-$ 3 and regions 8 $-$ 15), the power-law represents the contribution from unresolved point sources (which account for 29\% and 33\% of the total X-ray flux in the northern and southern outflows respectively), and in the central regions, it accounts for the AGN contribution. Regions 4 \& 5, 8 and 9 require a Gaussian ({\sc gauss}) component with a centroid fixed at 6.4 keV to model the contributions from the AGN's Fe K$\alpha$ line. Regions 4 \& 5 and 6 require an additional blackbody ({\sc bbody}) to model the soft excess from the AGN, and Region 7 necessitates an additional thermal plasma component to model the hot starburst produced gas. The models of Regions 6, and 7 include a {\sc PEXMON} component to account for the Compton reflection of the AGN continuum \citep{Nandra2007} as well as model the Fe K$\alpha$ line. The two regions along with region 4 \& 5 also include a third {\sc PHABS} component to account for AGN related absorption. The final models are listed in Table~\ref{table:models}

\begin{deluxetable*}{ll}
\tablecolumns{2}
\tablewidth{0pt} \tablecaption{Spectral Models\label{table:models}} 
\tablehead{\colhead{Reg.} & \colhead{Model} }  
\startdata
1-3, 10-15 & \textsc{const * phabs * phabs * (apec+powerlaw)}\\
4 \& 5 & \textsc{const * phabs * phabs * (apec+ bbody + phabs * (powerlaw + gauss))}\\
6 & \textsc{const * phabs * phabs * (apec+ bbody + phabs * powerlaw + pexmon)}\\
7 & \textsc{const * phabs * phabs * (apec+apec + phabs * powerlaw + pexmon)}\\
8, 9 & \textsc{const * phabs * phabs * (apec+powerlaw+gauss)} \\
\enddata

\end{deluxetable*}

To reliably constrain abundances, about 5000 counts are needed. Thus, to get near this threshold, we made two composite regions, one in the northern and one in the southern outflow. The southern composite region has 4000 counts and constrains abundances to a factor of two; the northern composite region has 11,400 counts and constrains abundances to 50\% of the best-fit values.Regions 1 $-$ 5 are combined into the southern composite region, and the same procedure was done in the north for regions 8 $-$ 15. Due to the complex nature of the central regions, we refrain from letting abundances vary there due to a large number of free parameters. We assume the central metallicity is solar based on previous work studying NGC~4945's central super star clusters that power the galactic wind \citep{Emig_2020}. The composite southern region has a model of \textsc{const*phabs*phabs*(vapec+powerlaw+gauss)}, and the composite northern region has a model of \textsc{const*phabs*phabs*(vapec+vacx+powerlaw+gauss)}. 

In addition to the components mentioned above, we tested whether the AtomDB charge-exchange (CX) model component ({\sc vacx}) is necessary to account for the line emission. CX emission is produced when ions capture electrons from neutral material \citep{AtomDB}, and it has been found to contribute substantially to soft X-rays in starburst-driven outflows \citep{CX,Lopez2020,Lopez2023}. The inclusion of a CX component statistically improved the spectral fits of the composite northern region. However, the southern region did not need the ({\sc vacx}) component based on F-tests, likely because of the higher $N_{\rm H}^{\rm NGC4945}$ there. Thus, we did include an {\sc acx} component in the northern composite spectrum model but not in the southern model.

\section{Results}\label{sec:results}

Figure~\ref{fig:spectraplots} shows the extracted X-ray spectra from six of the 15 regions (two southern, two central, and two northern regions). In region 6 (the southern central region), we detect several emission lines from Mg, Si, S, Ar, Ca and Fe. Fe K$\alpha$ is found in four regions (5 $-$ 8) that are associated with the AGN. Fe {\sc xxv} (with a centroid energy of $\approx$6.67 keV) is only apparent in regions 6 and 7 due to the presence of a hotter phase in the starburst there. 

Using the models described in Section ~\ref{sec:methods}, we fit the spectra and found the best-fit values of the column density $N^{\rm NGC4945}_{\rm H}$, temperature $kT$, and metal abundances of the northern (regions 8 $-$ 15) and southern composite (regions 1 $-$ 5) regions (see Table~\ref{table:fitresults}). The fits yield $\chi^2$/d.o.f. of 531/531 for the north and 284/272 for the south. In Table~\ref{table:values} we show the best-fit $N^{\rm NGC4945}_{\rm H}$ and $kT$ for all 15 regions. Both $N^{\rm NGC4945}_{\rm H}$ and $kT$ peak in the center, with $N^{\rm NGC4945}_{\rm H}=(10.1_{-1.90}^{+1.99})\times10^{22}\:\mathrm{cm^{-2}}$ and $kT$=$1.04_{-0.18}^{+0.20}$ keV, and decrease with distance along the outflows in both directions. On average, the southern outflow has higher column densities than the northern outflow due to NGC~4945's disk being along its line of sight, leading to larger uncertainties to the hot gas properties there. 

In Table~\ref{table:agnpar} we show the best fit parameters related to NGC~4945's AGN. These values are the absorbing column $\rm N_H^{AGN}$, absorbed powerlaw slope $\Gamma_{\rm pl}$, and the equivalent widths of the Fe K$\alpha$ line. 

\begin{deluxetable*}{lcccccccccr}
\tablecolumns{11}
\tablewidth{0pt} \tablecaption{Spectral Fit Results\tablenotemark{a} \label{table:fitresults}} 
\tablehead{\colhead{Reg.} & \colhead{$r$} & \colhead{$N_{\rm H}^{\rm NGC4945}$} & \colhead{$kT$} &  \colhead{O/O$_{\sun}$} & \colhead{Ne/Ne$_{\sun}$} & \colhead{Mg/Mg$_{\sun}$} & \colhead{Si/Si$_{\sun}$} & \colhead{S/S$_{\sun}$} & \colhead{Fe/Fe$_{\sun}$} & \colhead{$\chi^{2}$/d.o.f.} \\
\colhead{} & \colhead{(kpc)} & \colhead{($\times10^{22}$~cm$^{-2}$)} & \colhead{(keV)} & \colhead{} & \colhead{} & \colhead{} & \colhead{} & \colhead{} & \colhead{}
}  
\startdata
North & $+$0.25 & $0.16_{-0.08}^{+0.09}$ & 0.68$\pm0.04$ & $<0.37$ & $0.81_{-0.38}^{+0.71}$ & $0.87_{-0.30}^{+0.53}$ & $1.09_{-0.32}^{+0.56}$ & $1.41_{-0.69}^{+0.97}$ & $0.22_{-0.06}^{+0.11}$ & $531/531$\\
South & $-$0.17 & $0.64_{-0.34}^{+0.11}$ & $0.34_{-0.06}^{+0.15}$ & $0.51_{-0.24}^{+2.20}$ & $0.28_{-0.11}^{+0.41}$ & $0.16_{-0.14}^{+0.34}$ & 1 & 1 & $0.15_{-0.05}^{+0.14}$ & $284/272$\\
\enddata
\tablenotetext{a}{Abundances with values of 1 are frozen to solar values in the fits.}
\end{deluxetable*}

From the best-fit values, we calculate other properties of the outflow for each region, such as the electron number density $n_{\rm e}$, the thermal pressure $P/k$, and the cooling time $t_{\rm cool}$. The electron number density $n_{\rm e}$ of the thermal plasma is estimated using the best-fit normalizations $norm$ of the {\sc apec} components, where $norm = (10^{-14}EM)/4\pi D^2$ and the emission measure is $EM = \int n_{\rm e} n_{\rm H} dV$. We set $n_{\rm e}=1.2n_{\rm H}$, integrate over the volume $V$, and calculate $n_{\rm e}$ given $n_{\rm e} = (1.5\times10^{15}normD^2/fV)^{1/2}$, where $f$ is the filling factor that we assume is $f=1$. 

To compute $V$ for each region, we assume that each region has a cylindrical volume of radius $R$ that is estimated based on the broadband X-ray (0.5 $-$ 0.7 keV) surface-brightness profiles along the outflows' major axis. $R$ is defined as the scale encompassing 68\% of the X-ray surface brightness; this percentile captures most of the observed outflow emission and traces structure well. 

The thermal pressure is computed as $P/k=2n_{\rm e}T$, and the cooling time is $t_{\rm cool} = 3kT/\Lambda n_{\rm e}$. $\Lambda$ is the radiative cooling function (in units of erg$^{-1}$ cm$^{-3}$) calculated using CHIANTI ~\citep{CHIANTI}, assuming a thin, thermal plasma at solar metallicity.

The resulting $n_{\rm e}$, $P/k$, and $t_{\rm cool}$ for all 15 regions are listed in Table \ref{table:values}. The thermal pressure peaks in the center, specifically in region 7 with value $P/k= 1.9\times10^{7}$~ K~cm$^{-3}$, with elevated pressures in regions 6 ($P/k=2.1\times10^{7}$ K~cm$^{-3}$) and 8 ($P/k= 1.0\times10^{7}$ K~cm$^{-3}$) as well. In the southern regions, the values vary from (1 $-$ 2.6)$\times10^{6}$ K~cm$^{-3}$, and in the northern regions they vary from (0.06$-$2.9)$\times10^{6}$ K~cm$^{-3}$, decreasing with distance from the starburst disk. The longest cooling time is in region 4 \& 5, with $t_{\rm cool}$= 26.1~Myr. Region 11 has the shortest cooling times of $t_{\rm cool}$= 2.24 Myr.

\begin{deluxetable*}{lcccccccccc} 
\tablecolumns{7}
\tablewidth{0pt} \tablecaption{Physical Parameters of the NGC~4945 Disk Center and Outflow Regions \label{table:values}} 
\tablehead{\colhead{Reg.} & \colhead{Distance}& \colhead{$N^{\rm NGC4945}_{\rm H}$} & \colhead{$kT$} &\colhead{$norm$} & \colhead{$R$} & \colhead{$V$} & \colhead{$n_{\rm e}$} & \colhead{$P/k$} & \colhead{$t_{\rm cool}$} & \colhead{$\chi^{2}$/d.o.f.} \\
\colhead{} & \colhead{(kpc)} &\colhead{($\mathrm{\times 10^{22}\:cm^{-2}}$)} & \colhead{(keV)}& \colhead{($\mathrm{cm^{-5}}$)} & \colhead{($\times10^{20}$ cm)} & \colhead{($\times10^{62}$~cm$^{3}$)} & \colhead{(cm$^{-3}$)}  & \colhead{($\times10^{6}$ K~cm$^{-3}$)} & \colhead{(Myr)}}  
\startdata
1 & $-$0.55 & $1.23_{-0.43}^{+0.40}$ & $0.64_{-0.32}^{+0.35}$ & 5.27$\times10^{-5}$ & 8.90 & 6.90 & 0.12 & 1.8 & 25.5 & 20/21 \\
2 & $-$0.46 & $1.29_{-0.31}^{+0.35}$ & $0.32_{-0.10}^{+0.17}$ & 2.90$\times10^{-4}$ & 7.26 & 4.59 & 0.35 & 2.6 & 4.19 & 59/53 \\
3 & $-$0.36 & $1.24_{-0.41}^{+0.30}$ & $0.32_{-0.09}^{+0.35}$ & 3.06$\times10^{-4}$ & 8.08 & 5.69 & 0.33 & 2.4  & 4.45 & 76/66 \\
4 \& 5\tablenotemark{a} & $-$0.21 & $2.56_{-0.73}^{+1.01}$ & $0.81_{-0.14}^{+0.19}$ & 1.29$\times10^{-4}$ & 7.94 & 11.3 & 0.15 & 1.4 & 26.1 & 145/129\\
6 & $-$0.06 & $10.1_{-1.90}^{+1.99}$ & $1.04_{-0.18}^{+0.20}$ & 3.05$\times10^{-4}$ & 1.51 & 0.19 & 1.74 & 21 & 3.30 & 637/468 \\
7\tablenotemark{b} & $+$0.03 & $3.78_{-0.33}^{+1.08}$ & $0.88_{-0.12}^{+0.22}$ & 6.14$\times10^{-5}$ & 1.64 & 0.24 & 0.72 & 7.4 & 6.15 & 679/526 \\
8 & $+$0.13 & $1.50_{-0.08}^{+0.08}$ & $0.67_{-0.05}^{+0.07}$ & 1.37$\times10^{-4}$ & 2.74 & 0.65 & 0.64 & 10 & 4.90 & 234/201 \\
9 & $+$0.23 & $0.77_{-0.10}^{+0.10}$ & $0.55_{-0.09}^{+0.07}$ & 6.05$\times10^{-5}$ & 5.20 & 2.36 & 0.22 & 2.9 & 11.8 & 111/121 \\
10 & $+$0.33 & $0.75_{-0.10}^{+0.15}$ & $0.49_{-0.13}^{+0.08}$ & 3.94$\times10^{-5}$ & 5.75 & 2.88 & 0.16 & 1.9 & 15.4 & 128/105\\
11 & $+$0.42 & $1.07_{-0.19}^{+0.13}$ & $0.22_{-0.03}^{+0.06}$ & 1.21$\times10^{-4}$ & 6.43 & 3.62 & 0.26 & 1.3 & 2.24 & 83/89 \\
12 & $+$0.52 & $0.96_{-0.18}^{+0.15}$ & $0.24_{-0.04}^{+0.06}$ & 7.29$\times10^{-5}$ & 6.84 & 4.09 & 0.19 & 1.1 & 3.93 & 98/70 \\
13 & $+$0.62 & $1.13_{-0.19}^{+0.15}$ & $0.22_{-0.04}^{+0.07}$ & 8.67$\times10^{-5}$ & 7.26 & 4.60 & 0.19 & 1.0 & 3.06 & 69/64 \\
14 & $+$0.72 & $0.79_{-0.27}^{+0.27}$ & $0.27_{-0.07}^{+0.10}$ & 2.84$\times10^{-5}$ & 9.58 & 8.02 &0.08 & 0.5 & 12.8 & 34/44 \\
15 & $+$0.82 & $1.26_{-0.29}^{+0.37}$ & $0.13_{-0.03}^{+0.06}$ & 7.03$\times10^{-6}$ & 10.4 & 9.45 & 0.04 & 0.06 & 4.72 & 28/29\\
\enddata
\tablenotetext{a}{Due to low signal in region 4 and 5, the regions were combined to better constrain the outflow properties.}
\tablenotetext{b}{This region required two thermal plasma components so the listed value is a flux-averaged value. The individual values are $kT_1=0.85_{-0.55}^{+0.11}$ keV and $kT_2=2.53_{-0.84}^{+0.85}$ keV.}
\end{deluxetable*}

\begin{deluxetable}{lcc}
\centering
\tablecolumns{3}
\tablewidth{0pt} \tablecaption{AGN Parameters\label{table:agnpar}}
\tablehead{\colhead{Parameter}  & \colhead{Reg. 6} & \colhead{Reg. 7}}
\startdata
$\rm N_H^{AGN}$ ($\times10^{22}$~cm$^{-2}$) & $383^{+173}_{-109}$ & $448^{+173}_{-110}$\\
$\rm \Gamma_{pl}$\tablenotemark{a}   & $1.9$ & $1.9$ \\
$\rm EW_{Fe\:K\alpha}$ (keV)  & $0.43^{+0.16}_{-0.17}$ & $0.88^{+0.20}_{-0.21}$ \\
\enddata  
\tablenotetext{a}{Frozen to 1.9 based on the constraints from \cite{Puccetti14}.}
\end{deluxetable}

\section{Discussion}\label{sec:discussion}

\subsection{Comparison to Previous Work on NGC~4945} \label{sec:previouswork_NGC4945}

Previous X-ray work on both the nucleus and outflows of NGC~4945 has been performed. A very similar analysis was conducted by \cite{Schurch2002} using 24 ks of {\it XMM-Newton} and 50 ks of {\it Chandra} X-ray observations (six times shallower than the data analyzed in this work) of NGC~4945 to study the galaxy's nucleus and surroundings within 1~kpc. They extracted a spectrum from a region of radius 3\arcsec\ centered on the AGN, and it contained $\approx$750 net counts that were fitted with a Compton-reflection component, an Fe K$\alpha$ line at 6.4~keV modeled by a Gaussian, and a hot thermal component. 

Additionally, they extracted a starburst plus outflow spectrum from an elliptical region that excluded the 3\arcsec\ nucleus region and contained $\approx$1600 net counts. The spectrum included emission lines from  K $\alpha$ neutral iron at 6.4 keV and helium-like iron at 6.7 keV. They fit the spectra with three thermal-plasma components with temperatures of $0.60\pm0.03$ keV, $0.87\pm0.08$ keV, and $6.0^{+1.1}_{-0.8}$ keV. We do not detect a third temperature component, our hottest component in region 7, $kT_2=2.53_{-0.84}^{+0.85}$~keV, is lower than their hottest component even within the uncertainties. Our northern and central region values are statistically consistent of the AGN and starburst spectral results of \cite{Schurch2002}. 

In \cite{Done2003}, an analysis of NGC~4945 was performed using the $\approx$50~ks of {\it Chandra} data as well as simultaneous Rossi X-ray Timing Explorer, {\it RXTE}, observations. Separate spectra were extracted from the nucleus and several starburst/superwind regions. The nuclear spectrum was a hard, broad-band continuum (which they modeled as a Compton-reflection continuum) and a Fe~K$\alpha$ emission line at 6.4 keV, consistent with our results. 

The spectra from the starburst regions had Fe K$\alpha$ emission that decreased with distance from the nucleus, and they modeled the data using a neutral reflection component as well as a thermal (\textsc{MEKAL}) plasma model. The best fits showed that column density and temperatures decreased going outward from the nucleus, comparable to the values presented in our work.

\cite{Marinucci2017} focused on the nucleus of the galaxy using 420 ks from five {\it Chandra} ACIS-S observations, including three observations used in this paper (ObsIDs 864, 14984, 14985) plus two with the High-Energy Transmission Grating Spectrometer (HETG spectrometer; ObsIDs 4899 and 4900). We did not analyze the HETG observations as the diffracted photons may blend with diffuse emission of the outflows. \cite{Marinucci2017} extracted spectra from five central circular regions, one in the nucleus and four along the circumnuclear region. The analysis focused on data $>$3 keV separated into three energy bands: 3 $-$ 6 keV, 6.2 $-$ 6.5 keV, and 6.6 $-$ 7.0 keV. The latter two bands were chosen to map the neutral Fe emission and the ionized Fe lines, respectively. The ionized iron was only found near the nucleus, while the neutral iron was present in both the nucleus and outer regions. This is consistent with our results where the Fe XXV line is observed in the central regions 6 and 7 while the Fe K$\alpha$ line is observed extending up to region 9. The northern-most region of \cite{Marinucci2017} included starburst-associated emission and required a thermal plasma of temperature $2.0^{+1.5}_{-1.0}$~keV, consistent with our analysis and previous results. We also find that the equivalent widths of the Fe K$\alpha$ line shown in Table~\ref{table:agnpar} are similar to those of \cite{Marinucci2017}.

\cite{Marinucci2017} also found a clump of gas with intense Fe {\sc xxv} emission $\approx$40~pc from the nucleus. This clump is roughly at the boundary between our regions 6 and 7 and may be one of the contributors to the Fe XXV line we observe in those regions. This finding showed that there is a complex morphology at the center of NGC~4945 and its surroundings. This may be due to the AGN affecting the nearby environment and may be why our spectral models varied substantially by region. 

\subsection{Comparison to M82 and NGC~253}
\label{sec:previouswork_starbursts}

\begin{figure}
    \centering    \includegraphics[width=\columnwidth]{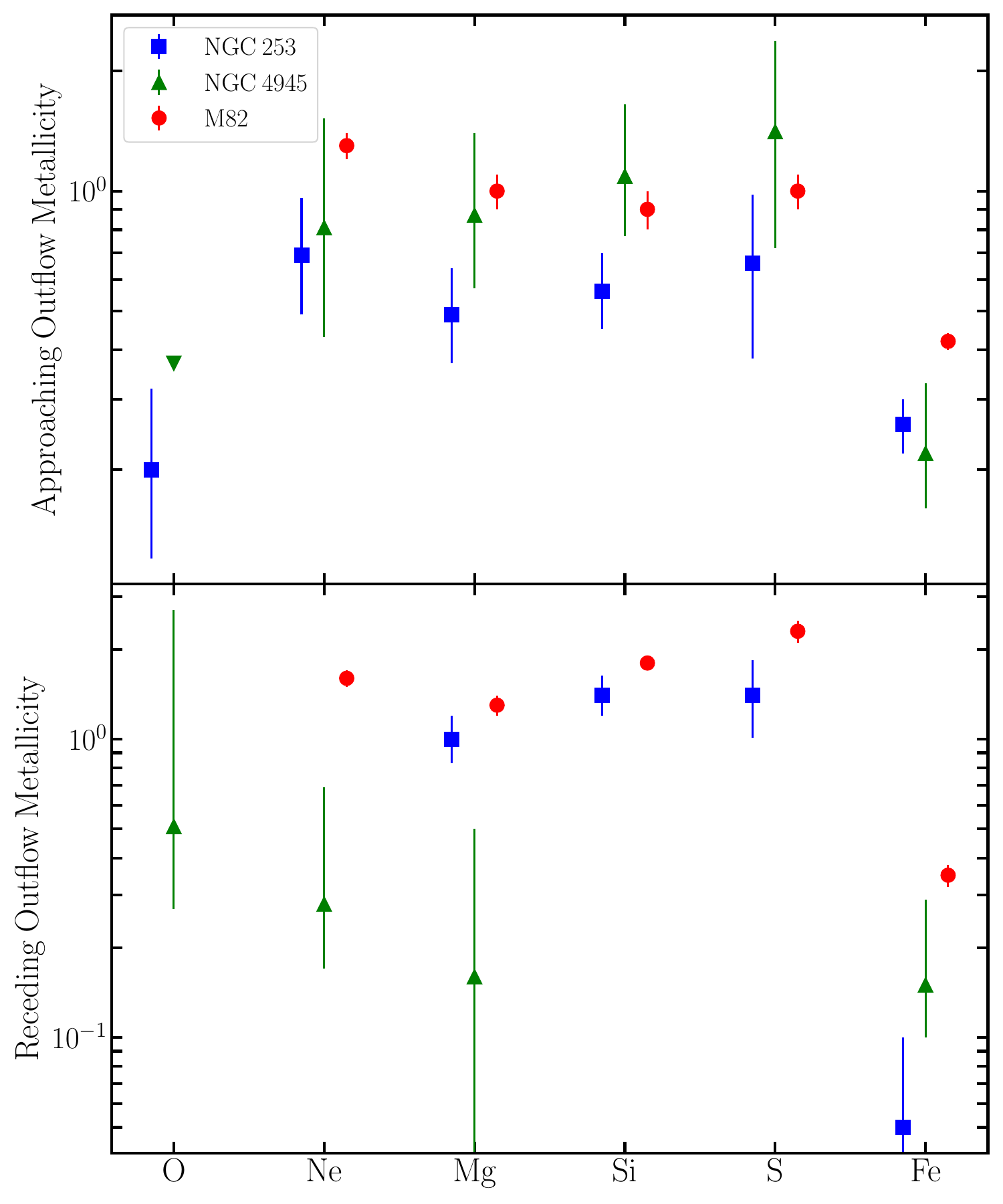}
    \caption{Metal abundances for NGC~4945, NGC~253 and M82 in green triangles, blue squares, and red dots, respectively, at a distance of $\sim$~0.2 kpc from the disk center used for NGC~4945.}
    \label{fig:metalcomp}
\end{figure}

The analysis in this paper follows that performed in \cite{Lopez2020} of M82 ($D = 3.6$~Mpc) as well as in \cite{Lopez2023} of NGC~253 ($D = 3.5$~Mpc), two other nearby galaxies with starburst-driven outflows. \cite{Lopez2023} showed that the M82 and NGC~253 hot gas outflows contain the same metals, but their distributions differ with distance from the galactic disks, with the former profiles being flat, while the latter are more centrally peaked.

NGC~4945 is a third example with starburst-driven galactic outflows. Unlike M82 and NGC~253, NGC~4945 has an AGN which precludes measurement of metal abundances associated with the central starburst. We did have sufficient signal to measure metal abundances in the composite northern and southern outflow  regions (Table~\ref{table:fitresults}), but we were not able to produce metal abundance profiles as in M82 and NGC~253. 

However, for comparison between these sources, in Figure~\ref{fig:metalcomp}, we plot the metal abundances of these hot gas outflows at the same distance from their galactic disks as measured in NGC~4945. We divide the figure between the approaching outflows (top) and receding outflows (bottom) according to the inclination of these systems. The approaching outflows in M82 and NGC~253 are the southern outflows, while for NGC~4945 it is the northern outflow. 

In the approaching outflows, the abundances of NGC~4945, NGC~253, and M82 are statistically consistent with each other where constrained. We find that M82 has higher abundances in the receding outflow for all measured elements (note: oxygen was not constrained in the M82 spectral fits because of the high column density). In the receding outflows, NGC~4945 has the lowest abundances except for Fe; however the large errors in NGC~4945 (particularly in the receding outflow where the column density is large, attenuating the soft X-ray emission) make comparisons difficult.  

The scatter in the receding outflows may be a result of the galactic disks inclination. In all three galaxies, the approaching outflows have lower column densities compared to the receding ones. This lack of intervening material allows for higher signal, enabling better constraints on abundances (see Section~\ref{specanalysis}). By contrast, the receding outflows have higher column densities, lowering the signal and leading to larger error bars. Consequently, while the receding outflow abundances vary between the galaxies, the large scatter may partly reflect the increased uncertainties in the measurements. 

The similarity between the abundances of the galaxies' approaching outflows may be the result of similar outflow driving mechanisms. All three galaxies have star-formation driven outflows powered by superstar clusters (SSCs) \citep{Maconi2000,Melo2005,Emig_2020,Mills2021,Levy2022}, and thus their chemical composition reflects the supernova enrichment. 

The similar hot gas metallicity profiles of the approaching outflows are noteworthy given the galaxies different stellar masses. NGC~253 and NGC~4945 have comparable stellar masses of $4.4\times10^{10}\:\mathrm{M_\odot}$ \citep{Bailin2011} and $3.8\times10^{10}\:\mathrm{M_\odot}$ \citep{Vulic2018} respectively, while M82 has a stellar mass of $10^{10}\:\mathrm{M_\odot}$ \citep{Greco2012}. It is observed that lower-mass galaxies have lower metallicity, possibly due to stronger feedback mechanisms ejecting metal rich gas more efficiently (the stellar-mass-metallicity relation: \citealt{Tremonti2004}). \cite{Chisholm2018} found that for nearby star-forming galaxies observed in the UV, the metal-loading factor decreases with stellar mass. It is not yet known whether this relationship holds for the hot phase, and it is possible that the lack of variation we find in the approaching outflows between these galaxies is because they do not span a large enough stellar mass range. In the future, X-ray studies of starburst-driven outflows for galaxies that span several dex in stellar mass would be valuable to explore metal loading across gas phases.

\begin{figure*}
    \centering   
    \includegraphics[width=\textwidth]{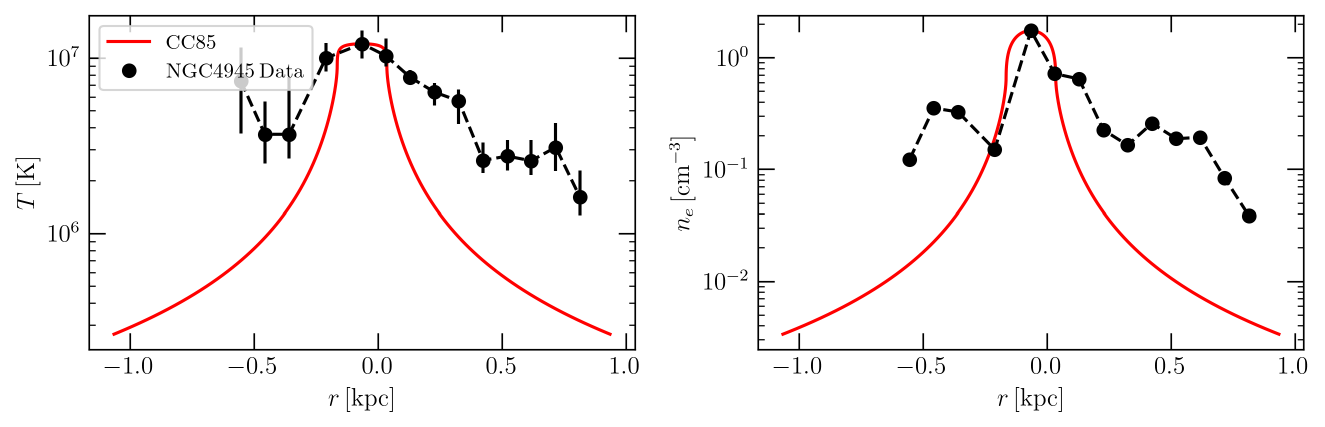}
    \caption{ {\it Left}: Best-fit hot gas temperatures $T$ values (black points) as a function of distance along the minor axis from the center of the galaxy (defined as $r = 0$). The red line shows the CC85 adiabatically expanding wind model $T$ prediction. The CC85 parameters used were $R=100$ pc, $\dot{M}_\mathrm{SFR}=4.17\,\mathrm{M_\odot \, yr^{-1}}$, $\alpha \simeq 0.12$ and $\beta \simeq 0.14$}. {\it Right}: Best-fit electron densities $n_{\rm e}$ (black points) compared to CC85 predictions. The measured temperature and density profiles are much broader than predicted, possibly due to mass loading or a non-spherical wind geometry.
    \label{fig:cc85}
\end{figure*}

\subsection{Comparison to Wind Models} \label{sec:windmodels}

The temperature $kT$ and density $n_{\rm e}$ profiles presented in \ref{sec:results} can be compared to galactic wind model predictions to constrain outflow properties. The CC85 model presented by \citep{CC85} describes a SN-powered galactic wind that assumes spherical symmetry and adiabatic expansion. This model is useful for comparison to observations because of its simplicity. 

In the CC85 model, the starburst radius ($R$) and the total energy ($\dot{E}_\mathrm{T}$) and mass-loading ($\dot{M}_\mathrm{T}$) rates are the only parameters. The energy injection rate into the region defined by $R$ is determined by the equation $\dot{E}_\mathrm{T} = \alpha \times 3.1 \times 10^{41} \times (\dot{M}_\mathrm{SFR}/M_\odot \, \mathrm{yr^{-1}) \, [ergs \, s^{-1}]}$, where  $\dot{M}_\mathrm{SFR}$ represents the star-formation rate, and $\alpha$ is a dimensionless parameter. The mass-loading rate is given by $\dot{M}_\mathrm{T} = \beta \times \dot{M}_\mathrm{SFR}$, where $\beta$ is another dimensionless parameter. Setting $R =100$~pc \citep{Maconi2000} and $\dot{M}_\mathrm{SFR}=4.17\,\mathrm{M_\odot \, yr^{-1}}$ (\citealt{Bendo2016}; adjusted to adopt our assumed distance to NGC~4945), $\alpha \simeq 0.12$ and $\beta \simeq 0.14$.

In Figure~\ref{fig:cc85}, we show the measured hot gas temperatures $T$ and electron densities $n_{\rm e}$ (black points) with the CC85 model predictions (red lines) for comparison. Both panels are plotted as a function of distance along the outflow, where $r = 0$ corresponds to the center of the galaxy. When comparing our data with the CC85 model predictions, we find divergence: the measured profiles are broader than those of the models at distances larger than the starburst region of 100~pc. 

This discrepancy between the CC85 models and observed hot gas profiles is similar to those found by \cite{Lopez2020} for M82 and by \cite{Lopez2023} for NGC~253. The inclusion of physical processes in the models like mass-loading and non-spherical wind geometry is necessary to produce profiles more consistent with observational results \citep{Nguyen2021}.

\subsection{Mass Outflow Rates} 
\label{sec:massoutflowrates}

As performed in \cite{Lopez2023}, using the estimated electron number density $n_{\rm e}$ and region volume $V$ (listed in Table~\ref{table:values}), the mass outflow rates can be calculated. For the southern outflow, we use measurements from regions $1-5$, and for the northern outflow, we use measurements from regions $8-15$. The mass of the outflows is calculated as $M_{\rm O} = n_e m_H V/ 1.2f^{1/2}$ where $V$ is the volume of a region and $f$ is the gas filling factor assumed to be 1. The mass outflow rate is then $\dot M = M_{\rm O} v/r$ where $v$ is the outflow velocity and $r$ is the vertical distance traveled. 

We find a mass of $3.5\times10^{5}\:M_{\odot}$ for the northern outflow and a mass of $4.2\times10^{5}\:M_{\odot}$ for southern outflow. For the mass outflow rates assuming a hot wind velocity of $v = 10^{3}$~km~s$^{-1}$, we find a rate of $0.52v_3\:M_{\odot}/\rm{yr}$ in the northern outflow and $1.1v_{3}\:M_{\odot}/\rm{yr}$ in the southern outflow, where $v_{3}$ is the outflow velocity in units of $10^{3}$~km~s$^{-1}$. With the assumptions made, the total hot phase outflow mass rate is $1.6\:M_{\odot}~\rm{yr}^{-1}$, less than the cold molecular mass outflow rate of $\sim20\:M_{\odot}~\rm{yr}^{-1}$ \citep{Bolatto2021} and comparable with the ionized phase of $1.6\:M_{\odot}~\rm{yr}^{-1}$ \citep{Heckman1990}. 

\subsection{Charge Exchange Contribution}
\label{sec:CX}

\begin{figure}
    \centering    \includegraphics[width=\columnwidth]{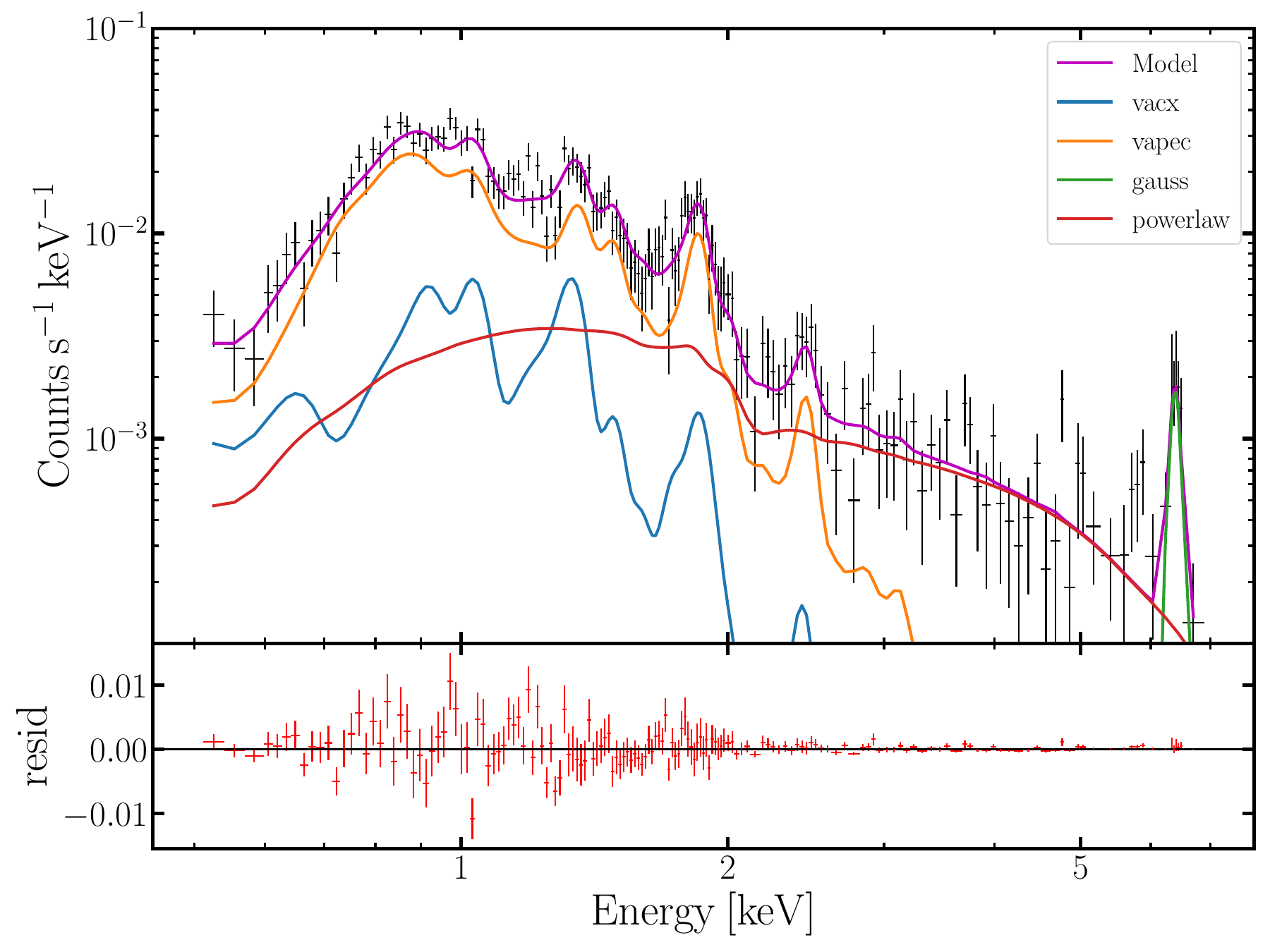}
    \caption{Spectrum from the composite northern region of NGC~4945 (regions 8-15)}, with the total spectral model, individual model components, and residuals plotted. The \textsc{vacx} component (the blue line) corresponds to the charge exchange emission and contributes 12\% to the total $0.5-7$~keV emitted flux.
    \label{fig:north_cx}
\end{figure}

As described in Section \ref{sec:methods}, we include a CX component in our spectral fits, and we find it contributes 12\% of the total broadband X-ray emission in regions $8-15$ of the northern outflow (see Figure~\ref{fig:north_cx}). CX was not necessary in the spectral models of the southern outflow, likely because of the higher intrinsic column density there obscuring the soft X-rays.

Charge exchange, the stripping of an electron from a neutral atom by an ion, contributes significantly to the soft X-ray emission of galactic outflows. \cite{Liu2012} showed that the CX contribution to the K$\alpha$ triplet of He-like O, Ne and Mg contributed 90\%, 50\% and 30\%, respectively, for several nearby starburst galaxies. Across different locations in the M82 outflows, \cite{Lopez2020} found that the CX component contributes up to 25\% of the broadband ($0.5- 7$~keV) flux. In the case of NGC~253, up to 42\% of the X-ray emission is produced by CX \citep{Lopez2023}. 

In our analysis of NGC~4945, we find that the CX component contributes 12\% to the broad-band flux, relatively small compared to the results from M82 and NGC~253. We note that this 12\% may be lower-limit on the CX contribution as the high intrinsic column density in NGC~4945 precludes detection of the soft emission where CX is most significant. Recent work by \cite{Okon2023} shows that CX emission in M82 may be concentrated at the interfaces of hot gas with swept-up cool clouds. If the CX arises similarly in the outflows of NGC~4945, then a lower CX flux contribution may suggest lower mass-loading and/or smaller cool clouds entrained in the hot superwind.

\section{Conclusions}\label{sec:conclusions}

We analyze six {\it Chandra} observations (totaling 330~ks) of NGC~4945 to obtain images and spectra of the hot galactic outflows extending 0.55 kpc south and 0.85 kpc north from the starburst nucleus where an AGN is present. 

We define 15 regions (two from the galactic disk/starburst and thirteen from the outflows; Figure~\ref{fig:spectraplots}) and model their spectra to find the best-fit parameters for the temperatures $kT$, intrinsic column densities $N^{\rm NGC4945}_{\rm H}$, and electron density $n_{\rm e}$ (Table~\ref{table:values}). The temperature and density profiles are broader than predicted from a spherically symmetric, adiabatically expanding wind (Figure~\ref{fig:cc85}). These results are consistent with recent work on M82 and NGC~253 and suggest the need to include additional physics in models, such as mass-loading and a non-spherical wind geometry. 

In addition to this analysis, we also model spectra from composite northern and southern regions of the outflow to derive metal abundances (Table~\ref{table:fitresults}). The abundances are greater in the northern outflow than in the southern outflow. The values are consistent with previous work on NGC~253 and M82's outflows (Figure~\ref{fig:metalcomp}). Further X-ray studies are necessary to explore the variation in hot gas metal loading with host galaxy properties.

The northern outflow X-ray spectra require a charge-exchange component which accounts for 12\% of the total broad-band, emitted X-ray flux (Figure~\ref{fig:north_cx}). This result adds to a growing body of literature indicating the need to include CX in X-ray spectral models of hot gas in starburst-driven winds.

We also provide an estimate of NGC4945's hot wind mass outflow rate: $1.6\:M_{\odot}~\rm{yr}^{-1}$ assuming a hot wind velocity of $10^3$ km~s$^{-1}$. We do not yet have dynamical information for the hot X-ray gas, but the superb spectral resolution of XRISM \citep{XRISM} will enable the measurement of hot wind kinematics. Velocity estimates are crucial to obtain better constraints on wind energy and on hot phase mass outflow rates.

\begin{acknowledgements}

NPB, SL, and LAL were supported by NASA's Astrophysics Data Analysis Program under grant number 80NSSC22K0496 and by the Heising-Simons Foundation through grant number 2022-3533. LAL also acknowledges the support of the Simons Foundation. We thank the OSU Galaxy/ISM Meeting for useful discussions. DDN acknowledges funding from NASA 21-ASTRO21-0174.
\end{acknowledgements}

\software{XSPEC \citep{XSPEC}, CIAO \citep{CIAO}, Astropy \citep{astropy}}

\bibliography{sample631}{}
\bibliographystyle{aasjournal}

\end{document}